\documentclass[twocolumn,showpacs,floatfix,superscriptaddress,amsmath,amssymb,prb]{revtex4}
\usepackage{mathrsfs}
\usepackage{txfonts}
\usepackage{amssymb}
\usepackage{graphicx}
\usepackage{hyperref}
\usepackage{ulem}
 \usepackage{overpic}
 \usepackage{psfrag}
\usepackage{tabularx}
\usepackage{array}
\usepackage{placeins}
\usepackage{subfigure}
\makeatletter
\renewcommand\subsection{\@startsection\texttt{}
{subsection}{1}{0mm}
 {-\baselineskip}
 {0.5\baselineskip}
{\FloatBarrier\normalfont\Large\bfseries}}
\makeatother
\newcommand{\be}{\begin{equation}}
\newcommand{\ee}{\end{equation}}

\newcommand{\PreserveBackslash}[1]{\let\temp=\\#1\let\\=\temp}

\begin{document}

\title{Universal Order Parameters and Quantum Phase Transitions: A Finite-Size Approach}
\author{Qian-Qian Shi}
\affiliation{College of Materials Science and Engineering,
Chongqing University, Chongqing 400044, The People's Republic of
China}
\affiliation{Centre for Modern Physics,
Chongqing University, Chongqing 400044, The People's Republic of
China}
\author{Huan-Qiang Zhou}
\affiliation{Centre for Modern Physics,
Chongqing University, Chongqing 400044, The People's Republic of
China}
\author{Murray T. Batchelor}
\affiliation{Centre for Modern Physics,
Chongqing University, Chongqing 400044, The People's Republic of
China}
\affiliation{Australian National University,
Canberra ACT 0200, Australia}

\begin{abstract}
We propose a method to construct universal order parameters for quantum phase transitions in many-body lattice systems.
The method exploits the $H$-orthogonality of a few near-degenerate lowest states of the Hamiltonian describing a
given finite-size system, which makes it possible to perform finite-size scaling and take full advantage of
currently available numerical algorithms. An explicit connection is established
between the fidelity per site between two $H$-orthogonal states and the energy gap between the ground state
and low-lying excited states in the finite-size system. The physical information encoded in this gap arising from
finite-size fluctuations clarifies the origin of the universal order parameter.
We demonstrate the procedure for the one-dimensional quantum formulation of the
$q$-state Potts model, for $q=2,3,4$ and 5, as prototypical examples, using finite-size data obtained from the density
matrix renormalization group (DMRG) algorithm.
\end{abstract}


\maketitle

Order parameters are pivotal to the Landau-Ginzburg-Wilson description of
phase transitions for a wide range of critical phenomena, both classical and quantum,
in many-body systems arising from spontaneous symmetry breaking (SSB).\cite{SSB,Anderson}
Despite their importance, relatively few systematic methods for determining order
parameters have been proposed.
One method proposed for quantum many-body lattice systems
utilizes reduced density matrices.\cite{FMO}
This approach takes advantage of the degenerate ground states (GSs) which appear
when a symmetry of the Hamiltonian is broken spontaneously in the thermodynamic limit.
An order parameter can be identified with an operator that distinguishes the degenerate GSs.
The idea of the method is to search for such an operator by comparing the reduced density matrices
of the degenerate GSs for various subareas of the system.
This method was demonstrated in models that are considered to exhibit dimer, scalar chiral, and topological orders.\cite{FMO}

Another approach makes use of the ground-state fidelity of a quantum many-body system.\cite{FID,LAD,KIT} 
For a quantum phase transition (QPT) arising from SSB,
a bifurcation appears in the ground-state fidelity per lattice site,
with a critical point identified as a bifurcation point.\cite{bif}
This in turn results in the concept of the universal order parameter (UOP),\cite{UOP}
in terms of the fidelity per site between a ground state and its symmetry-transformed counterpart.
The advantage of the UOP over local order parameters in characterizing QPTs
is that the UOP is model independent, and thus universal, in sharp contrast with
local order parameters, which are usually determined in an {\sl ad hoc} fashion.

UOPs have been calculated with tensor network (TN) algorithms for systems with translational invariance.
For Hamiltonians possessing symmetry group $G$ with $g$ the element of $G$, UOPs
for translational invariant infinite-size systems are defined based on the orthogonal degenerate
GSs corresponding to SSB, as a measure of distinguishability between ground state
$|\psi\rangle$ and quantum state $g|\psi\rangle$, which can be interpreted in terms of the
fidelity $F$ as a measure of the similarity between two states.\cite{NILSEN}

Such UOPs satisfy the basic definition of an order parameter: namely in the SSB phase,
with $|\psi\rangle$ and $g|\psi\rangle$ two of the degenerate GSs,
the corresponding UOP is nonzero, whilst in the symmetric phase, with $g|\psi\rangle \equiv |\psi\rangle$,
the UOP is zero.
It has been demonstrated that such UOPs can successfully describe the symmetry broken phases in both
one-dimensional and two-dimensional quantum systems.~\cite{UOP,UOP2D}

Since SSB occurs only in the thermodynamic limit, this construction of UOPs only makes sense in infinite-size quantum many-body systems. 
It is clearly desirable however, to construct UOPs directly from finite-size systems.
This will not only make it possible to perform finite-size scaling, but also make it possible to take full advantage of
currently available numerical algorithms, such as quantum Monte Carlo,\cite{MC} finite-size density matrix renormalization group (DMRG),\cite{DMRG} and
finite-size TN algorithms.\cite{TN}
Here we propose and test a specific scheme to do this in the finite-size context for systems with SSB.


{\sl Construction of UOPs from $H$-orthogonal states.--}
First, we recall the notion of fidelity per lattice site.
The fidelity $F(|\varphi_1\rangle,|\varphi_2\rangle)=|\langle\varphi_1|\varphi_2\rangle|$
between two states $|\varphi_1\rangle$ and $|\varphi_2\rangle$ scales as
$F(|\varphi_1\rangle,|\varphi_2\rangle)\sim d(|\varphi_1\rangle,|\varphi_2\rangle)^L$,
with $L$ the number of lattice sites.
The fidelity per lattice site~\cite{FID} $d$ is the scaling parameter
\begin{equation}
 \ln{d(|\varphi_1\rangle,|\varphi_2\rangle)}\equiv \lim_{L\rightarrow\infty}\frac{\ln F(|\varphi_1\rangle,|\varphi_2\rangle)}{L}, \label{fidelity}
\end{equation}
which is well defined in the thermodynamic limit.
With $|\varphi_1\rangle$ and $|\varphi_2\rangle$ ground states for different values of the control parameter,
the fidelity per lattice site is nothing but the partition function per site in the classical statistical lattice model.\cite{PAT}

We consider a hamiltonian $H$ of a quantum system possessing symmetry group $G$ with $g$
a unitary representation of $G$, i.e.,  $U^gH{U^g}^\dag=H$, with $U^g=g\otimes g\otimes g\ldots$
an infinite string of copies of matrix $g$.
With the SSB, the UOP is defined in terms of the fidelity per lattice site $d_{\infty}$ for an infinite-size system by\cite{note1}
\begin{equation}
{\cal O}=\sqrt{1-d_{\infty}^2} \, ,
 \label{Uorder0}
\end{equation}
where $d_{\infty}=|\langle\psi|g|\psi\rangle|^{1/L}$ with $L\rightarrow\infty$
the fidelity per lattice site between the ground state $|\psi\rangle$ and the quantum state $g|\psi\rangle$.\cite{UOP,UOP2D}

To study UOPs in the finite-size context, it is natural to think of using the fidelity per lattice site $d_L$
for systems of finite size $L$ to construct $d_{\infty}=\lim_{L\rightarrow\infty}{d_L}$.
However, applying the same definition of $d_L$ with the GSs of a finite-size system fails because
$d_{\infty}\equiv 0$ in all the range for $|\langle\psi|g|\psi\rangle|^{1/L}=0$ in both phases,
as SSB occurs only in an infinite-size system.
There is however, a way to overcome this obstacle for finite-size systems.

To outline the general idea, consider a system whose hamiltonian has $Z_{q}$, $q\in\mathbb{Z}^+$ symmetry.
At zero temperature, for the symmetry broken phase, we have $q$ degenerate ground states in the
thermodynamic limit and we do expect that the symmetry is spontaneously broken.
First we calculate $q$ low-lying states of this system with finite size $L$,
denoting the $i$th eigenstate and corresponding eigenvalue by $|\phi_{i}\rangle$ and $E_{i}$,
satisfying $H|\phi_{i}\rangle=E_{i}|\phi_{i}\rangle$.
The $Z_q$ symmetry can be understood as rotations among the variables pointing in the corresponding field directions.
Thus the Hilbert space associated with $Z_q$ can be separated into disjoint sectors labeled by the
phases $\omega_{m}=\exp{(2\pi i (m-1)/q)}$ with $m=1,2,\ldots,q$.
For our purpose, we construct $q$ $H$-orthogonal states $|\psi_{m}\rangle$ from the
$q$ low-lying states $|\phi_{m}\rangle$ by
 \begin{equation}
 |\psi_{m}\rangle=\sum_{j}{\omega_{m}^{j-1}c_{j} \, |\phi_{j}\rangle},
 \label{states}
 \end{equation}
in terms of the above defined phases $\omega_{m}$.

Here, each pair of the $q$ states are set to be orthogonal with respect to $H$, i.e.,
\begin{equation}
\langle\psi_{m}|H|\psi_{t}\rangle=0,
\label{Hcond}
\end{equation}
with $m \neq t$, so called $H$-orthogonality.\cite{note2}
The $q$ coefficients $c_{j}$ are fixed by the $H$-orthogonality and normalization conditions.
The fidelity per lattice site of two $H$-orthogonal states $|\psi_{t}\rangle$ and $|\psi_{m}\rangle$ takes the form
\begin{equation}
d_L=|\langle\psi_{m}|\psi_{t}\rangle|^{1/L}=\left| \sum_j{\omega_{t-m}^{j-1}|c_j|^2} \right|^{1/L}.
\label{dc}
\end{equation}
The final step in the scheme is to extrapolate the fidelity per lattice site $d_{L}$ between two
$H$-orthogonal states, $d_{\infty}=\lim_{L\rightarrow\infty}{d_L}$, with the
UOP following from the definition in Eq.~(\ref{Uorder0}).
This explains how degenerate GSs in the thermodynamic limit, responsible for symmetry breaking order, emerge from near degenerate low-lying states in the finite-size system.


{\sl Application: the $q$-state Potts model.--}
The quantum formulation of the $q$-state Potts model has hamiltonian\cite{qPotts}
\begin{equation}
\small H=-\sum_{i}{\left(\sum_{\alpha=1}^{q-1}{M_{i}^{\alpha}M_{i+1}^{q-\alpha}}+\lambda M_{i}^{z}\right)} \, ,
 \label{ham2}
\end{equation}
where $i$ are the lattice sites and $\lambda$ denotes the external
field along the $z$ direction.
The operators are written in matrix form:
\begin{equation}
M^1=\begin{bmatrix} 0 & I_{q-1} \\ 1 & 0
\end{bmatrix}, \quad
M^z=\begin{bmatrix} q-1 & 0 \\ 0 & -I_{q-1}
\end{bmatrix}
\end{equation}
with $M^i=(M^1)^i$ for $i=1,\ldots,q-1$, where $I_q$ is the $q\times q$ identity matrix.
The hamiltonian has $Z_{q}$ symmetry. For $\lambda<1$ the system is in the $Z_{q}$
symmetry broken ferromagnetic phase, and a symmetric
paramagnetic phase when $\lambda>1$.
It is well known that a continuous (discontinuous) QPT occurs
for $q\leq4$ ($q>4$) at $\lambda=1$ where the model is exactly solved.~\cite{baxter,hamer}

Consider first the case $q=2$, the quantum transverse Ising model, where matrices $M^1$ and $M^z$ are the
Pauli matrices $\sigma^x$ and $\sigma^z$.
Here the continuous QPT at $\lambda=1$ is between the $Z_{2}$ symmetry broken ferromagnetic phase and the symmetric paramagnetic phase.
We compute the ground state wave function $|\phi_{gs}\rangle$ and the first excited state wave function $|\phi_{ex1}\rangle$ for a system with finite size $L$, with corresponding ground state energy $E_{gs}$ and first excited state energy $E_{ex1}$.
Substituting $\omega_1=1$ and $\omega_2=-1$ into Eq.~(\ref{states}) gives the two $H$-orthogonal states
\begin{eqnarray}
|\psi_{1}\rangle&=&c_{1}|\phi_{gs}\rangle+c_{2}|\phi_{ex1}\rangle,\\
|\psi_{2}\rangle&=&c_{1}|\phi_{gs}\rangle-c_{2}|\phi_{ex1}\rangle,
\end{eqnarray}
which satisfy the $H$-orthogonality and normalization conditions
$\langle\psi_{1}|H|\psi_{2}\rangle=0$ and
$\langle\psi_{1}|\psi_{1}\rangle=\langle\psi_{2}|\psi_{2}\rangle=1$.
Thus, equivalently, we get
\begin{eqnarray}
|c_{1}|^2 E_{gs}-|c_{2}|^2 E_{ex1}=0,\\
|c_{1}|^2+|c_{2}|^2=1,
\label{condis}
\end{eqnarray}
with solution $|c_{1}|^2=E_{ex1}/(E_{gs}+E_{ex1})$ and
$| c_{2}|^2=E_{gs}/(E_{gs}+E_{ex1})$.
The fidelity per lattice site between the two $H$-orthogonal states is thus
\begin{equation}
d_{L}=\langle \psi_1|\psi_2\rangle|^{1/L}=\left| |c_{1}|^2-|c_{2}|^2 \right|^{1/L}
=\left|\frac{\delta_L}{E_{gs}+E_{ex1}}\right|^{1/L},
\label{gapfid}
\end{equation}
with energy gap $\delta_L=E_{ex1}-E_{gs}$.

\begin{figure}[t]
\centering
\includegraphics[width=0.48\textwidth]{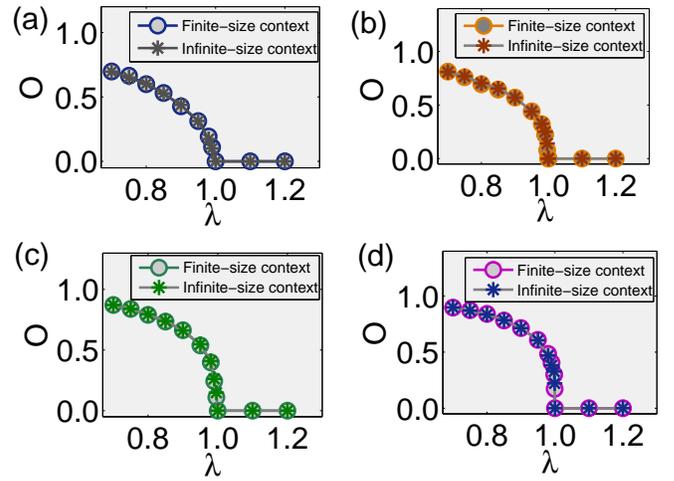}
\caption{(color online): Comparison of UOPs ${\cal O}$ for the $q$-state quantum Potts
model for $q=2,3,4$ and $5$ shown in $(a)$, $(b)$, $(c)$ and $(d)$, respectively.
In each case the UOP is calculated from finite-size systems and compared with the value obtained
in the infinite-size context.}
\label{order}
\end{figure}

In a similar fashion we have constructed the UOPs from the $q$ low-lying states of the
$q=3, 4$ and 5-state quantum Potts model.
The $q-1$ excited states share the same energy $E_{ex}$ above the ground state $E_{gs}$.
Proceeding as for the $q=2$ case, the coefficients $c_j$ in Eq.~(\ref{states}) ensuring the
$H$-orthogonality (Eq.~(\ref{Hcond})) and normalization conditions are obtained, with the expression for the fidelity per lattice site now
\begin{equation}
 d_{L}(\lambda)=\left|\frac{\delta_L(\lambda)}{(q-1)E_{gs}(\lambda)+E_{ex}(\lambda)}\right|^{1/L},
\label{dE}
\end{equation}
where $\delta_L(\lambda)=E_{ex}(\lambda)-E_{gs}(\lambda)$.
As such we have established an explicit connection between the fidelity per site between two
$H$-orthogonal states and the energy gap between the ground state and low-lying excited states,
which in turn renders clear physical implication for the UOP.
We emphasize that each pair of $H$-orthogonal states shares the same value of $d_L$ for given $\lambda$.

For values of the transverse field in the range $0.7 \le \lambda \le 1.3$,
we calculated the fidelity per lattice site $d_{L}(\lambda)$ between the $H$-orthogonal states
for finite-size systems $L$ ranging from 10 to 500 using
the DMRG algorithm.
We obtained $d_{\infty}(\lambda)$ and thus the UOP for each value of $\lambda$ by simple
extrapolation with $d_{L}(\lambda)=d_{\infty}(\lambda)+\alpha L^{-\beta}$.

Fig.~\ref{order} shows the UOPs obtained for $q=3, 4$ and 5
for values of the transverse field in the range $0.7 \le \lambda \le 1.3$
from finite-size systems $L$ ranging from 10 to 500 using the DMRG algorithm.
Also shown for comparison are the results obtained for infinite-size translation-invariant systems
with the infinite time-evolving block decimation (iTEBD) algorithm.\cite{itebd} 
The UOPs obtained from the finite-size approach outlined here and the
infinite-size approach match with a relative difference of less than $2.5$ percent,
which indicates the success of our scheme.
In general, as also shown in Fig.~\ref{order}, the UOP is seen to be capable of characterizing
the nature of the quantum phase transition.
For $q=2, 3$ and 4 there is a continuous phase transitions at $\lambda=1$, whilst for $q=5$
the first-order (discontinuous) phase transition can be seen at $\lambda=1$.
Here we remark that the fidelity per site has been demonstrated to be capable of detecting the discontinuous phase
transitions in this model through the so-called multiple bifurcation points.\cite{fidPotts}

\begin{figure}[t]
\vskip 2mm
\includegraphics[width=0.93\linewidth]{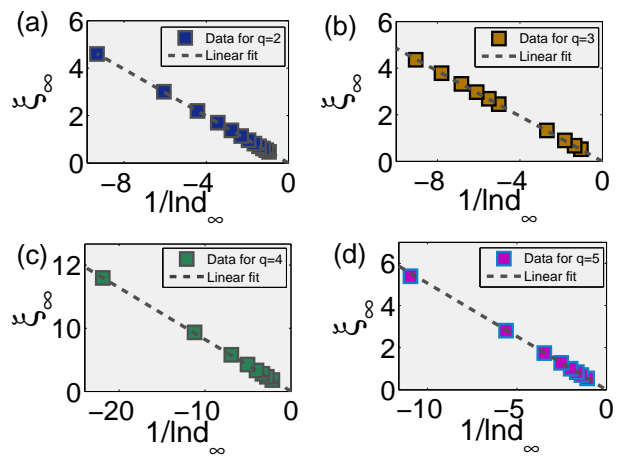}
\caption{The effective relation between the correlation length $\xi_{\infty}$ and the UOP.
In each case we calculate the correlation length $\xi_\infty(\lambda)$ and UOP ${\cal O}(\lambda)$
for control parameter $\lambda<1$ then fit $\xi_\infty(\lambda)$ and $\ln{d_\infty(\lambda)}$ to
the relation $\xi_\infty=-a/\ln{d_\infty}$, with $d_{\infty}(\lambda)=\sqrt{1-{\cal O}(\lambda)^2}$.
A simple linear fit gives the values (a) $a = -0.503$, (b) $a= -0.49$, (c) $a=-0.491$ and (d) $a= -0.506$.}\label{scale}
\end{figure}

{\sl Scaling.--}
For the $q$-state Potts model, the $q$ low-lying eigenstates are the single ground state and
$q-1$ degenerate first excited states.
The energy gap $\delta_L$ for a system of finite size $L$ obeys the relation
$\delta_L \sim d_{L}^L$ as Eq.~(\ref{dE}) indicates.
In the SSB phase with $\lambda<1$ away from the phase transition point,
the eigenspectrum is gapful and the energy gap $\delta_L$ is
related to the correlation length $\xi_{L}$ by $\delta_L \sim\exp{(-L/2\xi_{L})}$.
Taking $L\rightarrow\infty$, the fidelity per lattice site $d_\infty$ and correlation length $\xi_\infty$ are
expected to be related by
\begin{equation}
\xi_\infty=-\frac12 \frac{1}{\ln{d_\infty}}.
\label{corr}
\end{equation}
Fig.~\ref{scale} shows this expected relation between $d_\infty(\lambda)$ and $\xi_\infty(\lambda)$
for different values of $\lambda$.
Here, the data are mainly obtained using the iTEBD algorithm for infinite-size systems.
The results are consistent with the relation (\ref{corr}) holding throughout the SSB phase $\lambda < 1$.

\begin{figure}[t]
\vskip 4mm
\includegraphics[width=0.945\linewidth]{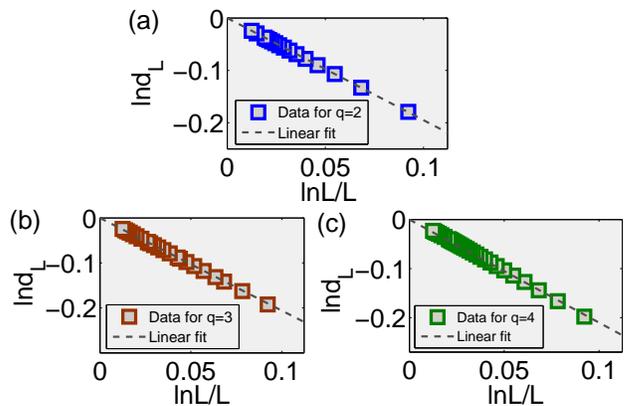}
\caption{Finite-size scaling of the fidelity per site $d_L$ at criticality.
In each case we fit $\ln{d_{L}}$ as a linear function of $\ln{L}/L$ and identify the
amplitude $b$ with data for system size $L$ ranging from $50$ to $500$.
The results are (a) $b=-1.96$, (b) $b=-2.06$ and (c) $b=-2.06$.}\label{scaleF}
\end{figure}

At the critical point $\lambda=1$, the correlation length $\xi$ and energy gap $\delta_L$
scale as $\xi\sim 1/\delta_L$.
With scale invariance at criticality, $\xi\sim L$, and thus $\delta_L\sim 1/L$.
Then with $d_{L}^L\sim \delta_L$ the expected relation between the fidelity per site
of the $H$-orthogonality states and finite size $L$ at criticality is $\ln{d_{L}}\sim -\ln{L}/L$.
The results presented in Fig.~\ref{scaleF} indicate that this relation is more precisely
\begin{equation}
\ln{d_{L}}\simeq -2\ln{L}/L.
\label{dL}
\end{equation}

\begin{figure}[t]
\includegraphics[width=0.93\linewidth]{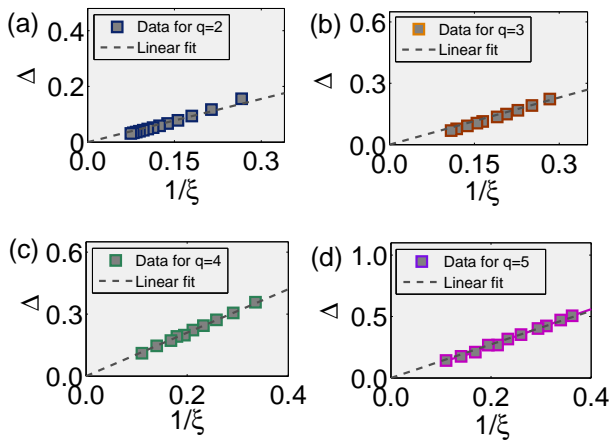}
\caption{Physical gap $\Delta$ vs correlation length $\xi$ for the $q$-state quantum Potts model at criticality.
For systems size $L$ ranging from 40 to 300, and a maximum number of 240 states kept during simulations
with the DMRG algorithm, we fit the data to $\Delta \, \xi = {\rm constant}$. For $q=5$ a finite gap
is obtained by extrapolating with finite truncation dimension from the iTEBD algorithm. In each case convergence is
expected towards the origin. However, at $q=5$ the mass gap terminates at the exact value $\Delta=0.0020544\ldots$.}
\label{gap}
\end{figure}

At the same time, keeping enough states with the DMRG algorithm,
we have accurately obtained the gap $\Delta$ between the ground state and the $(q+1)$-th lowest state
at criticality.\cite{note3}
Here it is known that $\Delta \, \xi = {\rm constant}$, which can be seen in the results of Fig.~\ref{gap}.
The case $q=5$ is particularly challenging because the mass gap is small, with the exact value
$\Delta=0.002 0544 \ldots$.\cite{hamer,gap}

{\sl Conclusions.---}
We have introduced a scheme for constructing UOPs to investigate QPTs using a set of $H$-orthogonal states
in finite-size systems.
We have established an explicit connection between the fidelity per site between two
$H$-orthogonal states and the energy gap between the ground state and low-lying excited states in the finite-size system,
which clarifies the physical meaning of the UOP.
This makes it possible to perform finite-size scaling and take full advantage of
currently available numerical algorithms.
The scheme has been tested for the $q-$state quantum Potts model
with $q=2, 3, 4$ and 5 using the finite-size DMRG algorithm.
We have demonstrated that the UOPs obtained
in the finite-size context agree with the UOPs obtained directly from the infinite-size context (Fig.~\ref{order}).
Our results suggest that, in the range where SSB occurs, the $H$-orthogonal states defined and
obtained in finite-size systems correspond to the $q$ degenerate ground states for the infinite system
when system size $L\rightarrow\infty$.
This clarifies how degenerate GSs emerge in the thermodynamic limit from low-lying near-degenerate states through $H$-orthogonality.
The UOPs we have thus defined are a further application of the fidelity per site in the characterisation of QPTs.

Furthermore, the general relation (\ref{corr}) between the correlation lengths and the fidelity is seen to hold in the SSB phase.
At criticality we have established the result (\ref{dL}) for the scaling of the fidelity per site.
Although we have considered UOPs from the point of view of finite-size systems
with $Z_q$ symmetry breaking,
it is anticipated that the scheme outlined here can also be extended and applied to
any system undergoing a phase transition characterized in terms of SSB.

This work is supported in part by the National Natural Science
Foundation of China (Grant Nos. 11174375 and 11374379) and by
Chongqing University Postgraduates' Science and Innovation
Fund (Project No. 200911C1A0060322).  M.T.B.  acknowledges support from the
1000 Talents Program of China.



\begin{thebibliography}{10}
\bibitem{note1} There are other possible definitions of the UOP. E.g.,
one could define ${\cal O} = 1 - d_\infty$ or ${\cal O} = \ln d_\infty$, which also vanish in the symmetric phase.
\bibitem{note3} Note that in principle one could perform calculations on the
equivalent staggered $XXZ$ Heisenberg chain, using the known mapping between the two models.\cite{hamer}
However, it is not clear how this mapping applies to the wavefunctions.
\end{thebibliography}
\end{document}